\documentclass[prd,preprint,superscriptaddress,a4paper,showpacs,showkeys,11pt,nofootinbib]{revtex4}

\usepackage{graphicx}
\usepackage{dcolumn}
\usepackage{bm}
\usepackage{multirow}
\usepackage{tabularx}
\usepackage{hyperref}
\usepackage{commath}

\begin{document}

\begin{flushright}
LU TP 16-XX\\
October 2016
\vskip1cm
\end{flushright}

\title{Exclusive vector meson photoproduction at the LHC and the FCC: A closer look on the  final state}


\author{G. Gil da Silveira}

\email[]{gustavo.silveira@cern.ch}

\affiliation{Instituto de F\'{\i}sica e Matem\'atica, Universidade Federal de Pelotas\\
Caixa Postal 354, CEP 96010-090, Pelotas, RS, Brazil}

\affiliation{Departamento de F\'{\i}sica Nuclear e de Altas Energias, Universidade do Estado do Rio de Janeiro\\
CEP 20550-013, Rio de Janeiro, RJ, Brazil}


\author{V. P. Gon\c calves}

\email[]{barros@ufpel.edu.br}

\affiliation{Instituto de F\'{\i}sica e Matem\'atica, Universidade Federal de
Pelotas\\
Caixa Postal 354, CEP 96010-090, Pelotas, RS, Brazil}

\affiliation{Department of Astronomy and Theoretical Physics, Lund University, 223-62 Lund, Sweden}



\author{M. M. Jaime}

\email[]{miguel.medina.jaime@cern.ch}

\affiliation{Instituto de F\'{\i}sica e Matem\'atica, Universidade Federal de Pelotas\\
Caixa Postal 354, CEP 96010-090, Pelotas, RS, Brazil}

\affiliation{Departamento de F\'{\i}sica Nuclear e de Altas Energias, Universidade do Estado do Rio de Janeiro\\
CEP 20550-013, Rio de Janeiro, RJ, Brazil}

\begin{abstract}
Over the past years the LHC experiments have reported experimental evidences for processes associated to  photon-photon and photon-hadron interactions, showing their potential to investigate the production of low- and high-mass systems in exclusive events. In the particular case of the photoproduction of vector mesons, the experimental study of this final state is expected to shed light on the description of the QCD dynamics at small values of the Bjorken-$x$ variable.  
In this paper we extend previous studies for the exclusive $J/\Psi$ and $\Upsilon$ photoproduction in $pp$ collisions based on the nonlinear QCD dynamics by performing a detailed study of the final state distributions  that can be measured experimentally at the LHC and at the Future Circular Collider. Predictions for the rapidity and transverse momentum distributions of the vector mesons and of final-state dimuons are presented for $pp$ collisions at $\sqrt{s} =$ 7, 13, and 100 TeV.     
\end{abstract}




\maketitle


\section{Introduction}

The exclusive production via photon interactions has attracted great interest given the experimental evidences reported by the Large Hadron Collider (LHC) experiments at CERN over the past years. The CMS experiment has reported results on the exclusive production of dileptons in $pp$ collisions at 7~TeV \cite{fwd-10-005,fwd-11-004}, the two-photon production of $W^{+}W^{-}$ pairs in $pp$ collisions at 7 and 8~TeV \cite{fsq-12-010,fsq-13-008}, and the $\Upsilon$ photoproduction in $p$-Pb collisions at 5.02~TeV \cite{fsq-13-009}. The other LHC experiments have reported similar results, e.g. the exclusive production of dileptons and $W^{+}W^{-}$ pairs by the ATLAS experiment in $pp$ collisions at 7 and 8~TeV \cite{atlas1,atlas2}, the $\Upsilon$ photoproduction in $pp$ collisions at 7 and 8~TeV by the LHCb experiment \cite{lhcb1,lhcb2,lhcb3}, and the $J/\Psi$ photoproduction in $p$-Pb collisions at 5.02~TeV and in Pb-Pb collisions at 2.76~TeV by the ALICE experiment \cite{alice1,alice2,alice3}. These results demonstrate the capability of the LHC experiments in probing the kinematic region of photon interactions and exclusive photoproduction at high energies, extending the previous results obtained at the Tevatron \cite{tevatron}, HERA \cite{zeus1,h11,zeus2,h12,zeus3,zeus4,h14}, and RHIC \cite{phenix} colliders. These set of measurements are the most comprehensive study of photon physics in hadronic collisions up to date, allowing more detailed studies of QED and the electroweak sector of the Standard Model and its extensions (for a recent review see, e.g. Ref.~\cite{review_forward}).

One of the main motivations for the study of the photoproduction of vector mesons in hadronic colliders is the possibility of probing the gluon distribution at small values of the Bjorken-$x$ variable \cite{vicber}, as well as the QCD dynamics at high energies \cite{vicmag}.
During the last decade, these ideas have been widely discussed in the literature considering different theoretical approaches that assume distinct underlying assumptions (see, e.g. Refs.~\cite{vicmag_outros,antoni,bautista}). Currently, we have that the experimental data are quite well described by models that consider the dipole approach and take into account nonlinear (saturation) effects in the QCD dynamics \cite{vicbruno_prd,bruno,griep,santos}. One important aspect of the colour dipole models is that the LHC predictions are parameter free, since their main elements -- the dipole-proton scattering amplitude and vector meson wave function -- have been constrained by the HERA data. 
Moreover, the energy dependence of the cross sections -- accessible at higher energies in the LHC than at HERA -- is determined by the QCD dynamics including nonlinear effects. The contribution of these effects increases with the  photon-proton centre-of-mass (c.m.) energy ($W_{\gamma p}$), being dependent on the mass of the vector meson. In particular, we expect a larger contribution for the $J/\Psi$ than for the $\Upsilon$ production. On the other hand, the recent LHCb data have been used to extract the gluon distribution ($xg$) assuming the validity of  the linear DGLAP evolution equation \cite{dglap} in the kinematical range considered \cite{jones,werner}, with the resulting gluon distribution being used as input to predict the cross sections for larger energies. It is important to emphasise that the $xg$ obtained through this procedure differs from that obtained in usual global analysis and also that the contribution of the next-to-leading order corrections for the exclusive vector meson photoproduction in collinear factorisation still are subjects of intense debate \cite{jones2}. Very recently, the BFKL approach have been applied in Ref.~\cite{bautista}  for the vector meson production in $pp$ collisions. These authors have obtained that the linear BFKL evolution \cite{bfkl} is capable to describe the energy dependence of the cross sections in the current energy range probed by the LHCb data if a fit of the initial transverse momentum profile of the proton impact factor is performed. The studies performed in Refs.~\cite{vicbruno_prd,bruno,jones,bautista} demonstrated that a clear determination of the underlying QCD dynamics is still not feasible by the simple comparison of the predictions with the experimental data for the rapidity distributions of the $pp$ cross sections and/or for the unfolded $\gamma p \rightarrow V p$ cross section. Consequently, the complementary study of other final state distributions is important to shed light on this subject.
In this paper we analyse in detail the photoproduction of vector mesons ($J/\Psi$ and $\Upsilon$) in $pp$ collisions at LHC energies whereas for the kinematical range expected to be probed by the Future Circular Collider (FCC) \cite{Arkani-Hamed:2015vfh}. We will focus in the final-state kinematics of the vector mesons decaying into muons and we will investigate the impact in the observables of two different assumptions for the energy dependence of the $\gamma p \rightarrow V p$ cross section. In particular, we  estimate the rapidity and transverse momentum distributions of the vector mesons, as well as the acoplanarity and transverse momentum balance  distributions of the dimuons from the decay of the vector mesons. In our study we will use the \texttt{SuperCHIC} v2.0 Monte Carlo (MC) event generator \cite{sc2}, modified to include the predictions of the nonlinear QCD dynamics for the energy dependence of the vector meson photoproduction cross sections derived in Refs.~\cite{vicbruno_prd,bruno}. Finally, it is important to emphasise that we will present the predictions for the photoproduction of vector mesons at the FCC for the first time in the literature.

This paper is organised as follows. In the next Section we present a brief review of the photoproduction of vector mesons in proton-proton collisions and discuss the approaches used in our analysis to estimate the $\gamma p \rightarrow V p $ ($V$ = $J/\Psi$ and $\Upsilon$) cross sections. 
Moreover, we present some details regarding the Monte Carlo event generator used in our study. In  Section~\ref{sec:res} we  present our predictions for the total cross sections and different final state distributions that can be analysed in the Run-II of the LHC as well as  at FCC. Finally,  in Section~\ref{sec:sum} we summarise the main conclusions of this work. 


\section{Photoproduction of Vector Mesons in Proton - Proton Collisions}
\label{sec:theory}

The exclusive photoproduction is taken as the diffractive interaction between an emitted photon from one of the colliding protons and the second proton. The photon emission is described in the equivalent photon approximation \cite{Fermi,weizsacker,williams}, which provides  a photon flux based on the energy spectrum of the emitted photons. This  flux can be obtained in terms of the electric and magnetic form factor of the proton, given by \cite{epa}
\begin{eqnarray}
\frac{\dif n}{\dif\omega} = \frac{\alpha}{\pi}\frac{\dif Q^{2}}{Q^{2}\omega} \left[ \left( 1 - \frac{\omega}{\sqrt{s}} \right) \left( 1 - \frac{Q_{min}^{2}}{Q^{2}} \right) F_{E} + \frac{\omega^{2}}{2s} F_{M} \right],
\end{eqnarray}
with $\sqrt{s}$ being the $pp$ c.m.~collision energy, $\omega$ and $Q^{2}$ the photon energy and virtuality, respectively, and $F_{i}$ are the proton electromagnetic form factors
\begin{eqnarray}
F_{E}(Q^{2}) &=& \frac{4m_{p}^{2}G_{E}^{2}(Q^{2})+Q^{2}G_{M}^{2}(Q^{2})}{4m_{p}^{2}+Q^{2}}, \\
F_{M}(Q^{2}) &=& G_{M}^{2}(Q^{2}),
\end{eqnarray}
where $G_{i}$ are the Sachs form factors \cite{sachs1,sachs2}
\begin{eqnarray}
G_{E}^{2} = \frac{G_{M}^{2}(Q^{2})}{7.78} = \frac{1}{(1+Q^{2}/0.71\textrm{ GeV}^{2})^{4}},
\end{eqnarray}
with the minimum photon virtuality being given by \cite{epa}
\begin{eqnarray}
Q_{min}^{2} = \frac{\omega^{2}}{\sqrt{s}}\frac{m_{p}^{2}}{\sqrt{s}-\omega},
\end{eqnarray}
for $\sqrt{s}-\omega\gg m_{p}$. The total cross section is given by the convolution of the photon spectrum and the $\gamma p \rightarrow V p $  cross section, which  has to be evaluated with a minimum energy to produce the central system,  as follows
\begin{eqnarray}
\sigma_{pp\to p \otimes V \otimes p} = 2 \int \frac{\dif n}{\dif\omega} \, \hat{\sigma}_{\gamma p\to Vp}(\omega,Q^{2}) \, \dif\omega\,\,,
\end{eqnarray}
where $\otimes$ indicates the presence of a rapidity gap in the final state. In the case  of two proton beams, the total cross section has to evaluated with a multiplicative factor of 2 to take into account the possibility of a photon emission from both protons. The rapidity distribution of the vector meson is given by 
\begin{eqnarray}
\frac{\dif\sigma_{pp\to p \otimes V \otimes p}}{\dif Y} = \frac{\dif n (\omega_+)}{\dif\omega_+} \hat{\sigma}_{\gamma p\to Vp}(+Y) + \frac{\dif n(\omega_-)}{\dif\omega_-} \hat{\sigma}_{\gamma p\to Vp}(-Y),
\end{eqnarray}
with the photon energies being given by  $\omega_{\pm}=(M_{V}/2)e^{\pm Y}$. This formula accounts for the forward and backward emission for the production of a  vector meson  with  rapidity $Y$.

The main input in the calculations is the $\gamma p \rightarrow V p $  cross section, which can be modelled in terms of different basic quantities, depending on the approach used to describe the process. In the collinear formalism the cross section is proportional to the square of the gluon distribution at $x \approx m_V/W_{\gamma p}$ \cite{ryskin,brodsky,martin}, which satisfies the linear DGLAP equation. In the $k_T$-factorisation approach,  it is given in terms of the square of the unintegrated gluon distribution, which satisfies the linear BFKL \cite{bfkl} or its nonlinear generalisations. On the other hand, in the dipole approach it is proportional to  $[{\cal{N}}(x,r,b)]^2$, where ${\cal{N}}$ is the forward dipole - proton scattering amplitude, which describes the interaction of a $q\bar q$ dipole of size $r$ with the proton at an impact parameter $b$ and its evolution is given by the nonlinear Balitsky-Kovchegov (BK) equation \cite{bk} in the Colour Glass Condensate (CGC) formalism \cite{cgc}. One have that in all approaches the energy dependence of the $\gamma p \rightarrow V p $ cross section is strongly dependent on the description of the QCD dynamics at high energies. In particular, we expect that although their predictions can be similar in a limited energy range, their extrapolations for higher values become very distinct, such that the experimental analysis of the process can be used to discriminate between the approaches. Recent LHCb data have provided data in the range of $W_{\gamma p}<$~2~TeV, but the Run-II of the LHC is expected to probe values of order of 5~TeV, while the FCC (100~TeV) should probe $W_{\gamma p} \approx$~15~TeV. As discussed before, the current data for the rapidity distributions and/or unfolded $\gamma p \rightarrow V p $ cross section can be described by different approaches based on distinct assumptions for the QCD dynamics and it is not clear if the isolated analysis of this distribution could discriminate between the different models in the future. Therefore, it is important to investigate the behaviour of the cross sections in the kinematical range that will be probed in the Run-II of the LHC and in the FCC, with particular emphasis in complementary final state distributions that can be directly compared with the experimental data. In order to obtain some estimates of these distributions, in what follows we will consider a phenomenological approach for the process that describe the current experimental data and can be implemented in a MC event generator.

Phenomenologically, the current experimental data for vector meson photoproduction can be described with a power-law function of the $\gamma p$ cross section with $W_{\gamma p}$, like $\sigma\propto W_{\gamma p}^{\delta}$. In particular, in Ref.~\cite{sc2}, the following power-law fit is assumed
\begin{eqnarray}
\frac{\dif\sigma^{\gamma p\to Vp}}{\dif t} = N_V \, \left( \frac{W_{\gamma p}}{W_{0}} \right)^{\delta_{V}} \beta_{V} e^{-\beta_{V}|t|},
\label{eq1}
\end{eqnarray}
where $W_{0}=$~1~GeV  and $\delta_{V}$ gives the slope of the differential cross section in terms of $W_{\gamma p}$, which is obtained by \cite{motyka_watt}
\begin{eqnarray}
\delta_{V} = \left. \frac{\partial\ln\sigma_{\gamma p}}{\partial\ln W_{\gamma p}} \right|_{W_{\gamma p}=W_{0}}.
\end{eqnarray}
Additionally, the parameter $\beta_{V}$ in Eq. (\ref{eq1}) is the slope of the proton-Pomeron vertex, which is parametrized by a function based on the Regge theory 
\begin{eqnarray}
\beta_{V} = \beta_{0} + 4\alpha^{\prime} \log \left( \frac{W_{\gamma p}}{\textrm{90 GeV}} \right),
\label{bv}
\end{eqnarray}
where $\beta_0 = 4.6$ GeV$^{-2}$ and $\alpha^{\prime} = 0.2$ GeV$^{-2}$ is the slope of the Regge trajectory, which are assumed universal for the two mesons. In the case of the $J/\Psi$ production, the parameters $N_{\Psi}$ and $\delta_{\Psi}$, which provide the normalisation and the slope of the differential cross section, respectively, were extracted in Ref.~\cite{sc2}  from the power-law fit to the available data of the exclusive (or elastic) photoproduction of the $J/\psi$ meson at HERA, resulting in $N_{\Psi}=$~(3.97~$\pm$~0.05)~nb and $\delta_{\Psi}=$~0.67~$\pm$~0.03. The particular case of the $\Upsilon$(1S) meson photoproduction is less accurate in the HERA energy regime, providing only a few data points with large uncertainties. In Ref.~\cite{sc2}, the authors have included the recent LHCb data in the fitting and have obtained $N_{\Upsilon}=$~5.7~pb and $\delta_{\Upsilon}=$~0.7. These values largely differ from previous values derived fitting only the HERA data, which were $N_{\Upsilon}=$~0.12~pb and $\delta_{\Upsilon}=$~1.6. As indicated in \cite{sc2}, the resulting predictions for $\sqrt{s} =$~13~TeV contain a 50\% uncertainty due to the error in the extracted parameters of the power - law fit. 
Finally, it is important to emphasise that the \texttt{SuperCHIC2} assumes that the photoproduction cross section is given by the Eq.~(\ref{eq1})  to produce event samples which can analysed using the \texttt{ROOT} data analysis framework \cite{ROOT}, with the default values for $N_V$ and $\delta_V$ being $(N_{\Psi},\delta_{\Psi}) = (3.97 \,\,\mbox{nb}, 0.67)$ and $(N_{\Upsilon},\delta_{\Upsilon}) = (5.7 \,\, \mbox{pb}, 0.7)$.

The resulting values for $N_V$ and $\delta_V$ discussed above are effective values constrained by the current data and are valid for a limited kinematical range. In particular, they are not associated to a specific model of the QCD dynamics. As a consequence, predictions for a higher energy range, where new dynamical effects are expected to modify the energy dependence of the cross sections, should be considered as educated guess. In order to investigate the impact of the nonlinear effects on the energy dependence and compare its predictions with the approaches that are available at \texttt{SuperCHIC2}, we have fitted the predictions to the results evaluated in Refs.~\cite{vicbruno_prd,bruno} within the colour dipole framework \cite{nik} and 
using the bCGC model \cite{kmw,amir} for the dipole-proton scattering amplitude. Then, we obtain that the energy dependence can be described using Eq.~(\ref{eq1}) with the following set of parameters: $(N_{\Psi},\delta_{\Psi}) = (10.25 \,\,\mbox{nb}, 0.49)$ and $(N_{\Upsilon},\delta_{\Upsilon}) = (3.85 \,\, \mbox{pb}, 0.76)$. It allows to use \texttt{SuperCHIC2} to extend the analyses performed in Refs.~\cite{vicbruno_prd,bruno} and obtain, e.g. the kinematical distributions of the vector mesons decaying into $\mu^{+}\mu^{-}$, which have been not studied previously in the literature. 

\begin{center}
\begin{figure}[t]
\includegraphics[width=0.4\textwidth]{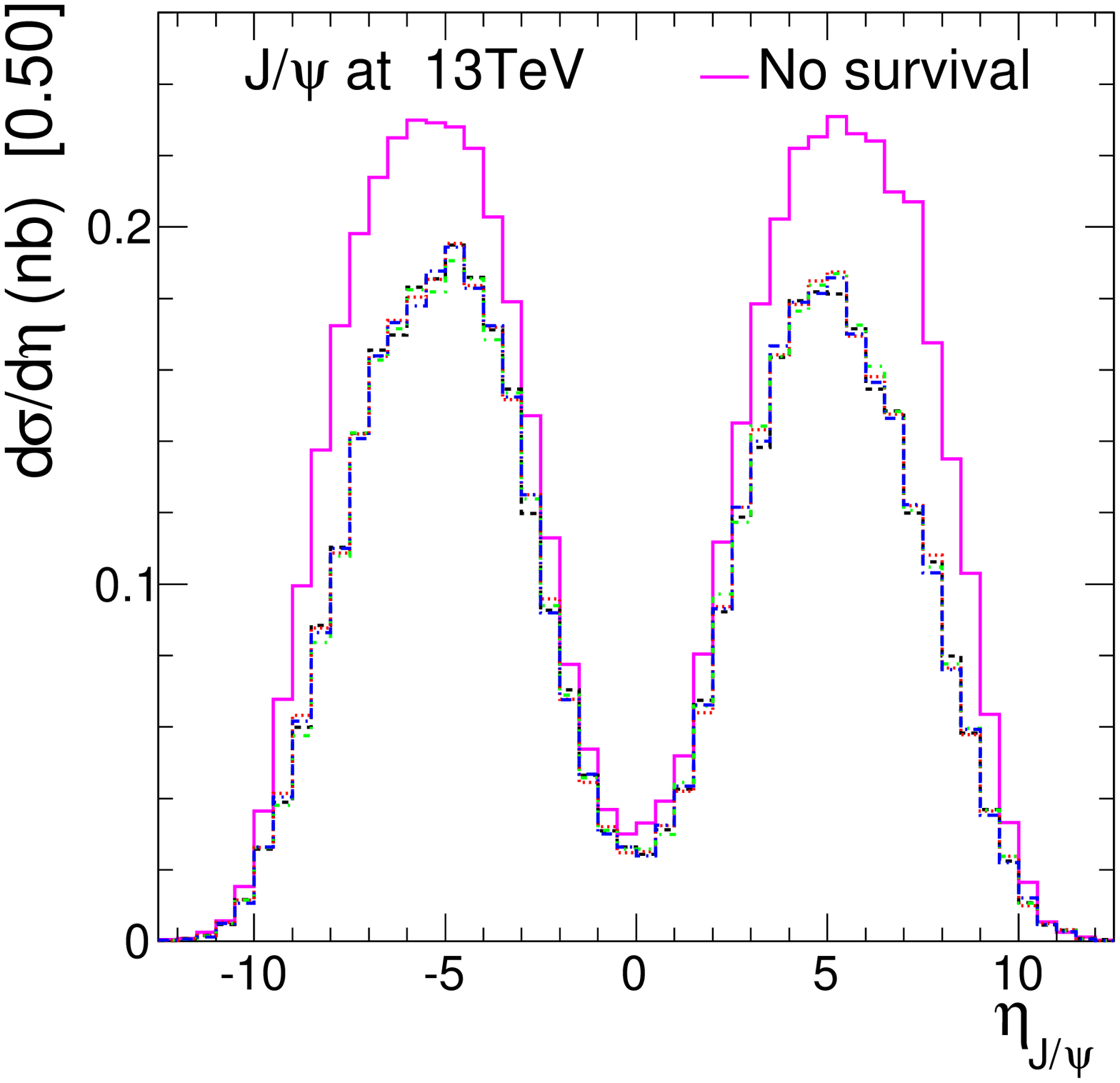}
\includegraphics[width=0.4\textwidth]{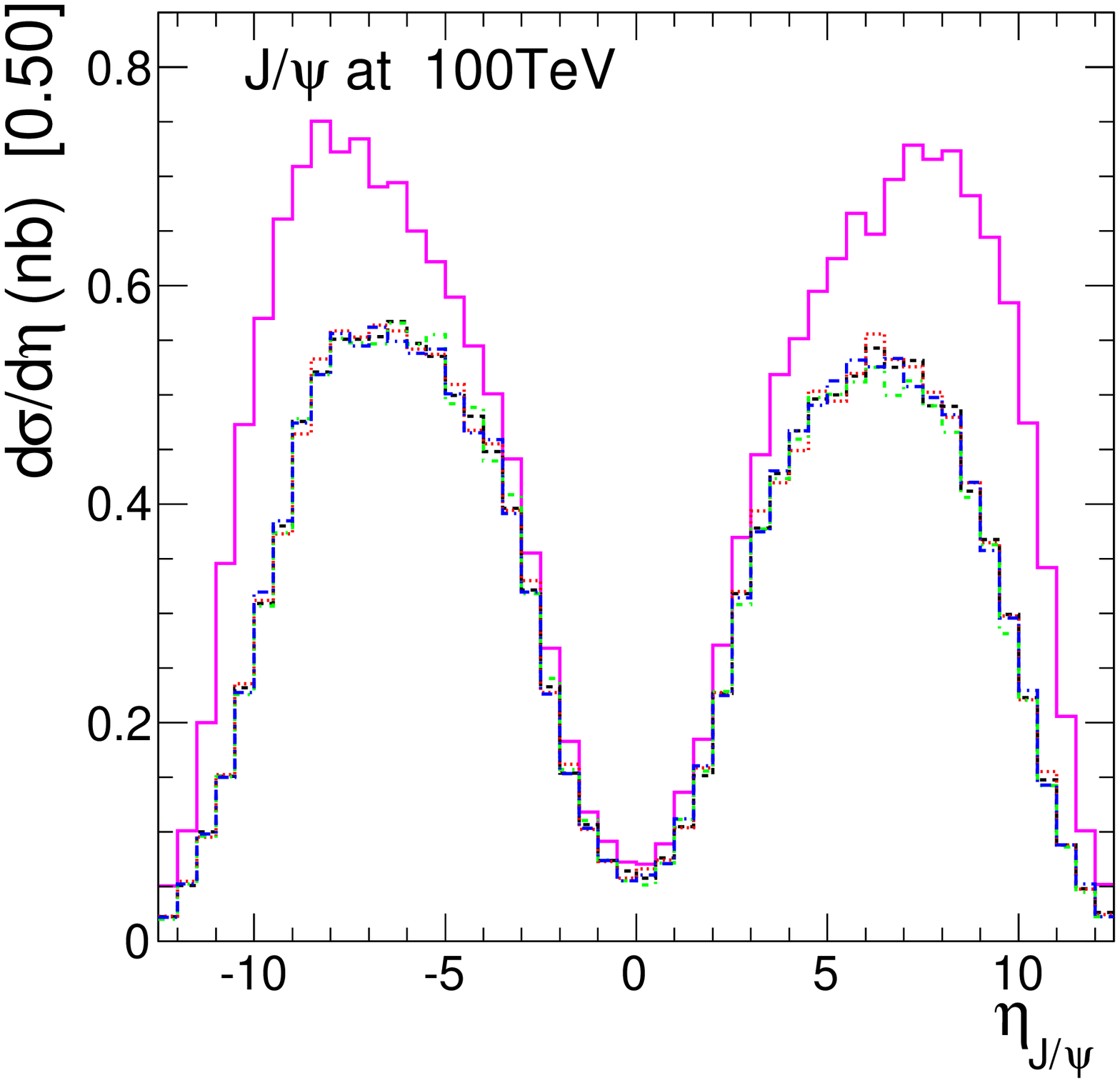}
\caption{\label{fig-sur}
Rapidity distribution for the exclusive $J/\Psi$ photoproduction  in $pp$ collisions at $\sqrt{s} =$~13 (left panel) and 100~TeV (right panel) considering different models for the survival probability. The prediction derived disregarding the extra soft reinteractions between the incident protons is represented by the solid line.}
\end{figure}
\end{center}


\section{Results}
\label{sec:res}

Our goal in this Section is to present our predictions for the exclusive vector meson photoproduction in $pp$ collisions at the LHC ($\sqrt{s} =$ 7 and 13~TeV)  and FCC ($\sqrt{s} =$~100~TeV) considering two different assumptions for the energy dependence of the $\gamma p \rightarrow V p$ cross section. However, before presenting the results, it is important to discuss the possible impact of soft interactions that can populate the rapidity gaps in the final state \cite{bjorken}. This subject have been intensively discussed in the last years \cite{kmr_gap,telaviv}, mainly motivated by the experimental data for dijet production in single diffraction events \cite{tevatron_sd,atlas}   that demonstrated that a gap survival factor should be taken into account in order to describe the data. The modelling and magnitude of this factor for the diffractive photon-hadron interaction still are themes of intense debate, with the possibility that it is equal to one being still valid. In particular,  the 
four different models presented in Ref.~\cite{kmr1} have been implemented in the \texttt{SuperCHIC2} in a fully differential format. Thus, the kinematical distributions can be properly estimated by including such effect, since the survival factor depends on the particular subprocess under consideration. In Fig.~\ref{fig-sur} we estimate the impact of the extra soft reinteractions  in the rapidity distributions for the $J/\Psi$ production at $\sqrt{s} =$~13 and 100~TeV using the default parameters for $N_{\Psi}$ and $\delta_{\Psi}$. We have that for central rapidities the prediction obtained disregarding the extra soft reinteractions, represented by the solid line, is very similar to those calculated using the four different models present in Ref.~\cite{sc2}. However, the survival corrections have a strong impact at large-$\eta$, reducing the cross sections by $\approx$35\%. We checked that similar suppressions are obtained using the bCGC model. As the treatment and magnitude of the survival probability remains a theme of intense debate, in what follows we will present our predictions assuming that it is equal to unity. 

\begin{center}
\begin{figure}[t!]
\hspace*{-3.5em}\includegraphics[width=0.55\textwidth]{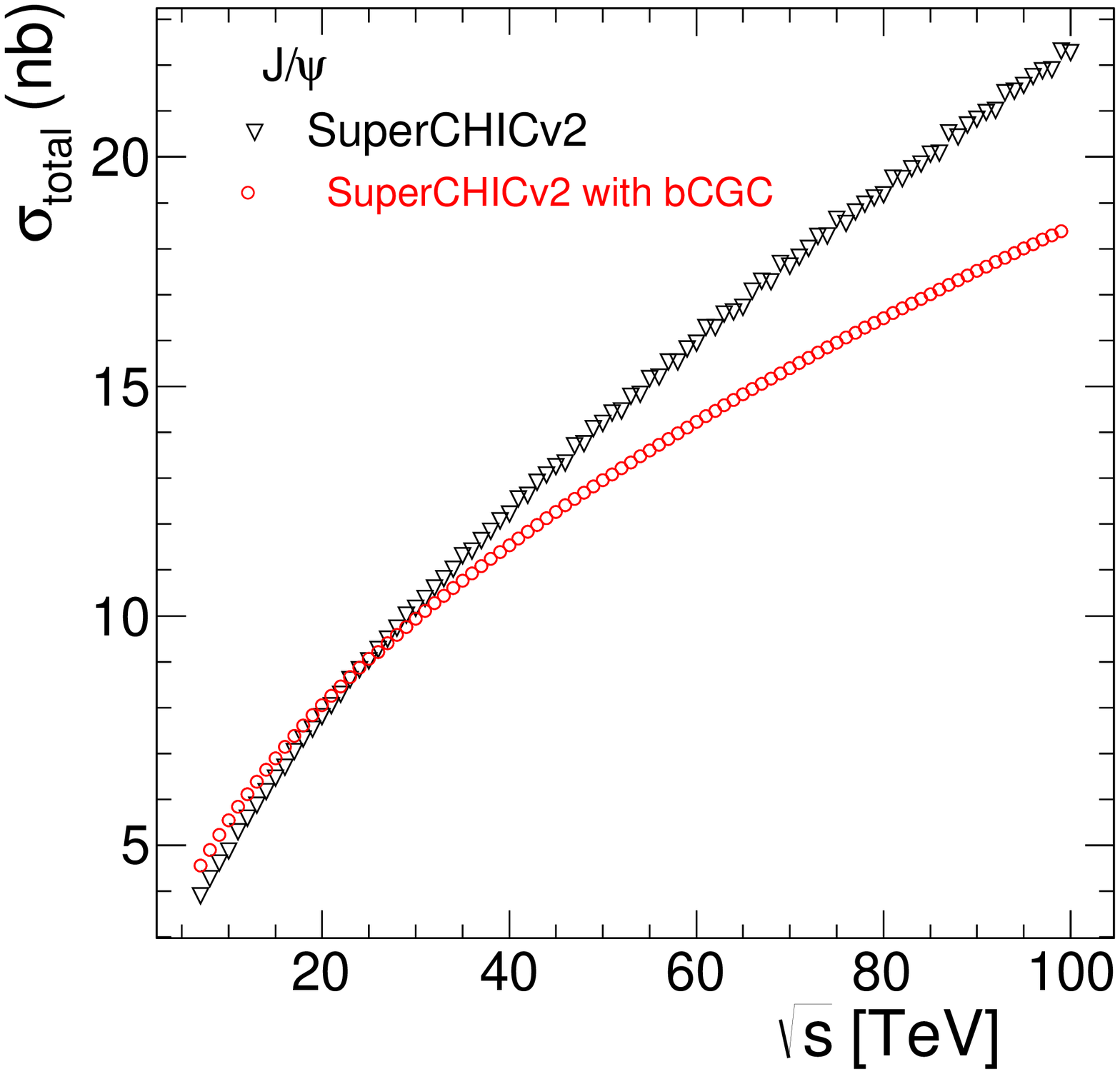}
\hspace*{-1.5em}\includegraphics[width=0.55\textwidth]{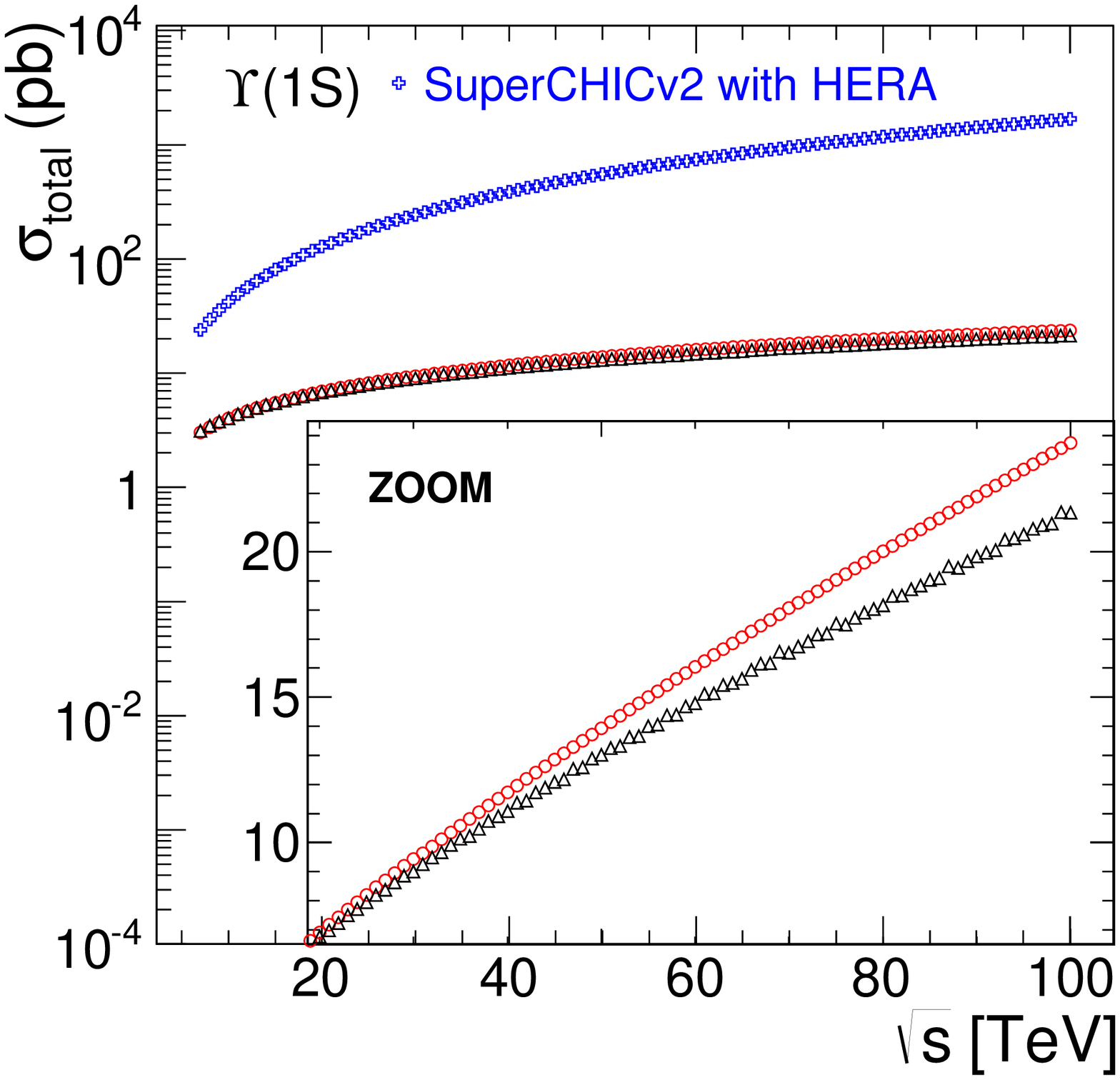}
\caption{\label{fig-E}
Predictions for the energy dependence of the total cross sections for the photoproduction of $J/\psi$ (left panel) and $\Upsilon$ (right panel) in $pp$ collisions considering the default values of the \texttt{SuperCHIC2} and those associated to the bCGC model. For comparison, we also present the predictions for $\Upsilon$ production obtained with the fitting parameters originally extracted from the HERA data.}
\end{figure}
\end{center}

In Fig.~\ref{fig-E} and Table~\ref{tab1} we present our predictions for the  energy dependence of the total cross sections for the photoproduction of $J/\Psi$ (left panel) and $\Upsilon$ (right panel) in $pp$ collisions considering the default values of the \texttt{SuperCHIC2} and those associated to the bCGC model. For comparison, we also present the predictions for $\Upsilon$ production obtained  with the fitting parameters originally extracted from the HERA data. For the $J/\Psi$ production, one have that the default and bCGC predictions are similar to LHC energies, which is expected since both models describe the current data. Moreover, they differ by $\approx$20\% at the FCC energy regime, with the bCGC model predicting smaller values. Such aspect is also expected, since at larger energies the contribution of the nonlinear effects increases, modifying the energy dependence of the $\gamma p$ cross section. In the case of the $\Upsilon$ production, one have that if the old parameters, derived fitting  the HERA data, are used in the calculations, the resulting predictions for the total cross section is one order of magnitude larger than those obtained with the parameters derived by the fitting of the HERA and LHCb data, simultaneously, that are the default parameters of the \texttt{SuperCHIC2}. Besides, we have that the default and bCGC predictions are similar at LHC energies and differ by $\approx$20\% at the FCC, with the bCGC one predicting larger values for the total cross section. As in the $J/\Psi$ case, we would expect the opposite result.
We believe that this difference is associated to the fact that the default values for $(N_{\Upsilon},\delta_{\Upsilon})$  have been obtained by fitting the current set data, which is smaller in comparison to the $J/\Psi$ one. As a consequence, these effective values have a large uncertainty. Surely, the Run-II data can improve this situation.

\begin{center}
\begin{table}[t!]
\begin{tabularx}{\textwidth}{@{}l *4{>{\centering\arraybackslash}X}@{}}
\hline\hline
\multirow{2}{*}{Process} & \multirow{2}{*}{$\sqrt{s}$ (TeV)} & \multicolumn{3}{c}{$\sigma[pp\to p+(\gamma p)\to p+V+p ]$} \\
        &             & HERA \cite{hera-fit} & Default \cite{sc2} & bCGC \cite{vicbruno_prd,bruno}              \\
\hline
\multirow{3}{*}{$J/\psi\to\mu^{+}\mu^{-}$}
        & 7.               & --             & 3.90~nb        & 4.58~nb           \\
        & 13.              & --             & 5.87~nb        & 6.43~nb           \\
        & 100.             & --             & 22.42~nb       & 18.65~nb          \\
\hline
\multirow{3}{*}{$\Upsilon\textrm{(1S)}\to\mu^{+}\mu^{-}$}
        & 7.               & 23.91~pb       & 3.13~pb        & 3.02~pb           \\
        & 13.              & 64.34~pb       & 4.95~pb        & 4.92~pb           \\
        & 100.             & 1682.72~pb     & 21.50~pb       & 23.91~pb          \\
\hline\hline
\end{tabularx}
\caption{\label{tab1}Exclusive vector meson photoproduction cross sections  in the $\mu^{+}\mu^{-}$ decay channels for $pp$ collision energies of the LHC and the FCC. }
\end{table}
\end{center}

\begin{center}
\begin{figure}[t!]
\includegraphics[width=0.45\textwidth]{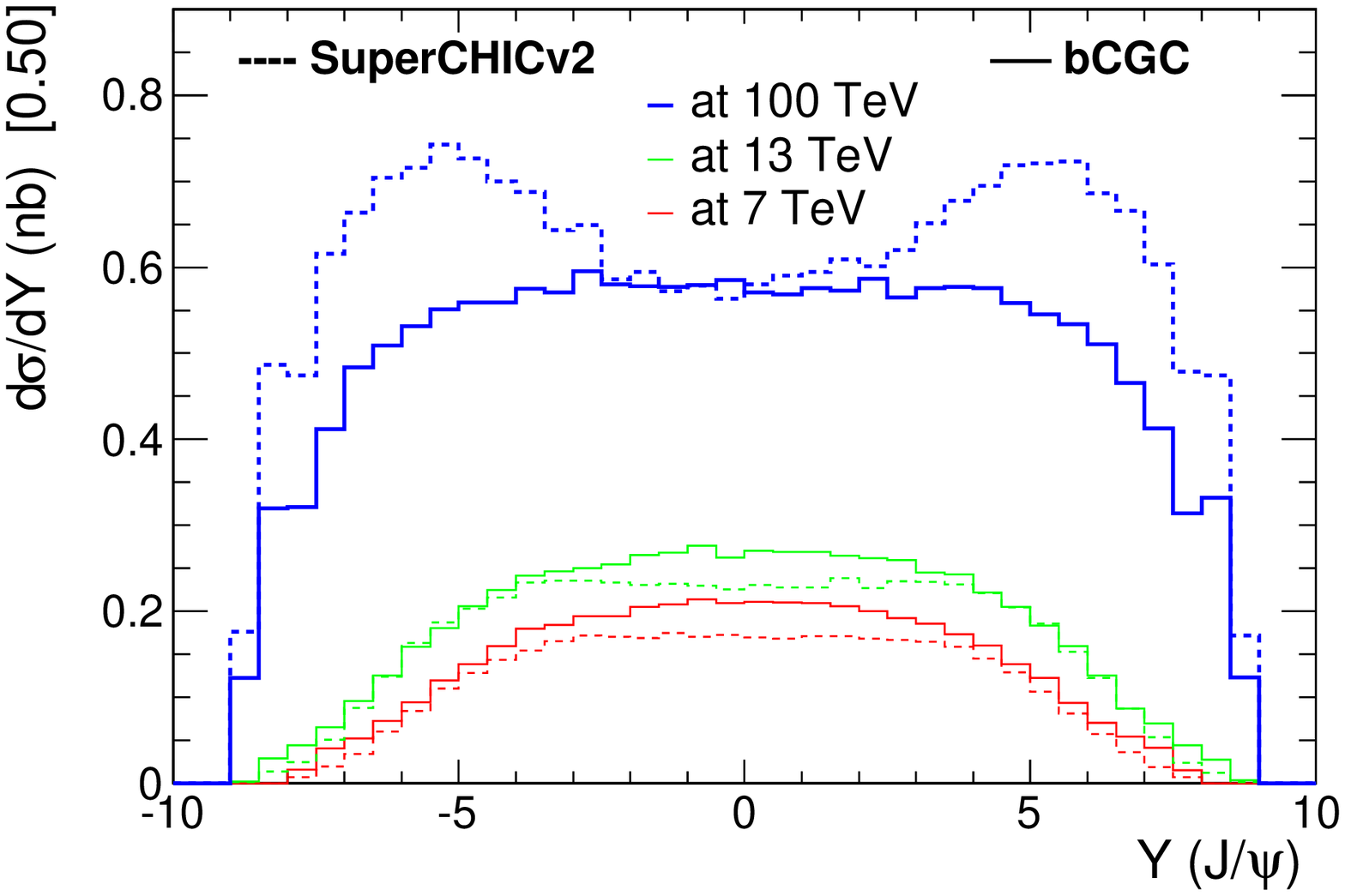}
\includegraphics[width=0.45\textwidth]{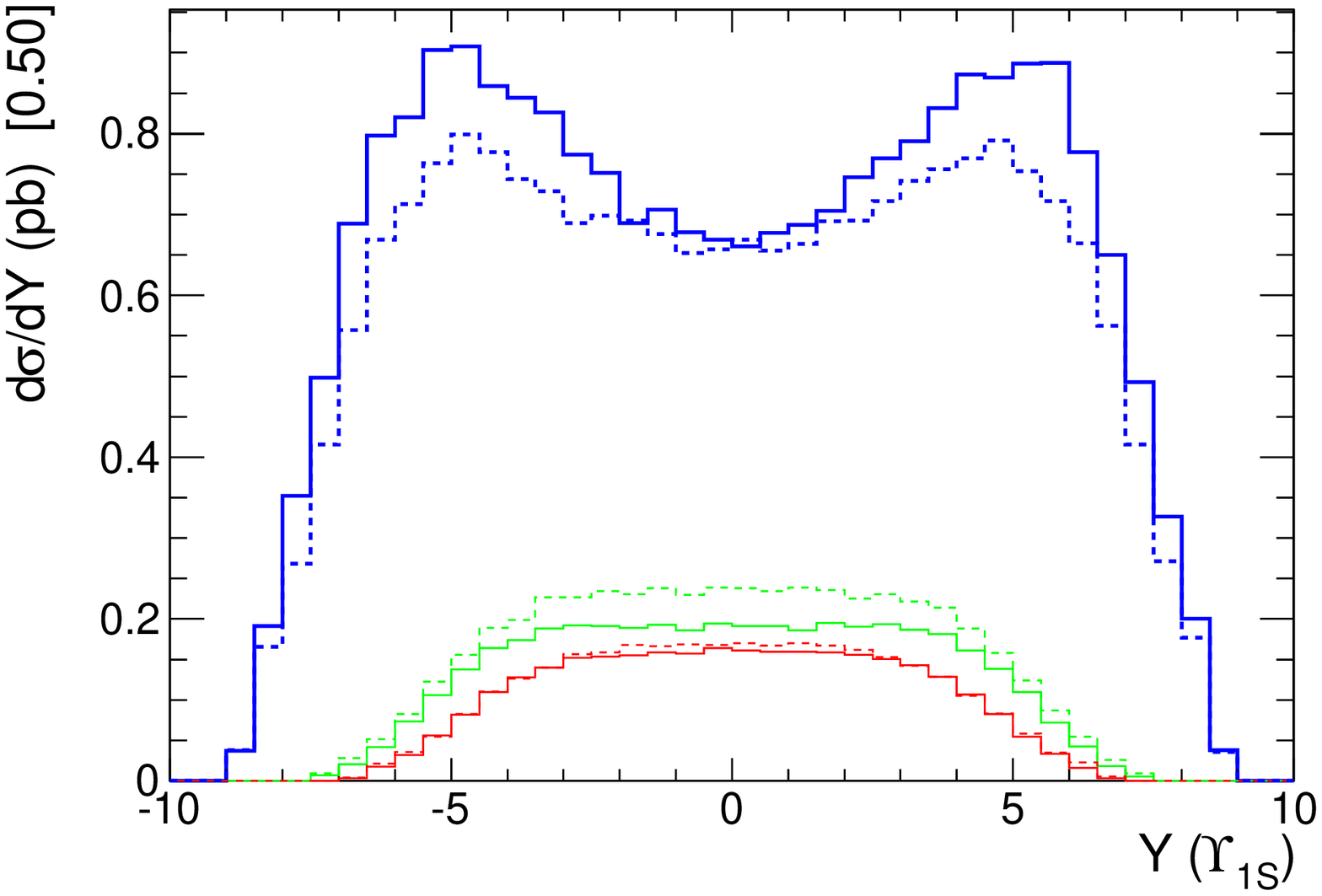}
\includegraphics[width=0.45\textwidth]{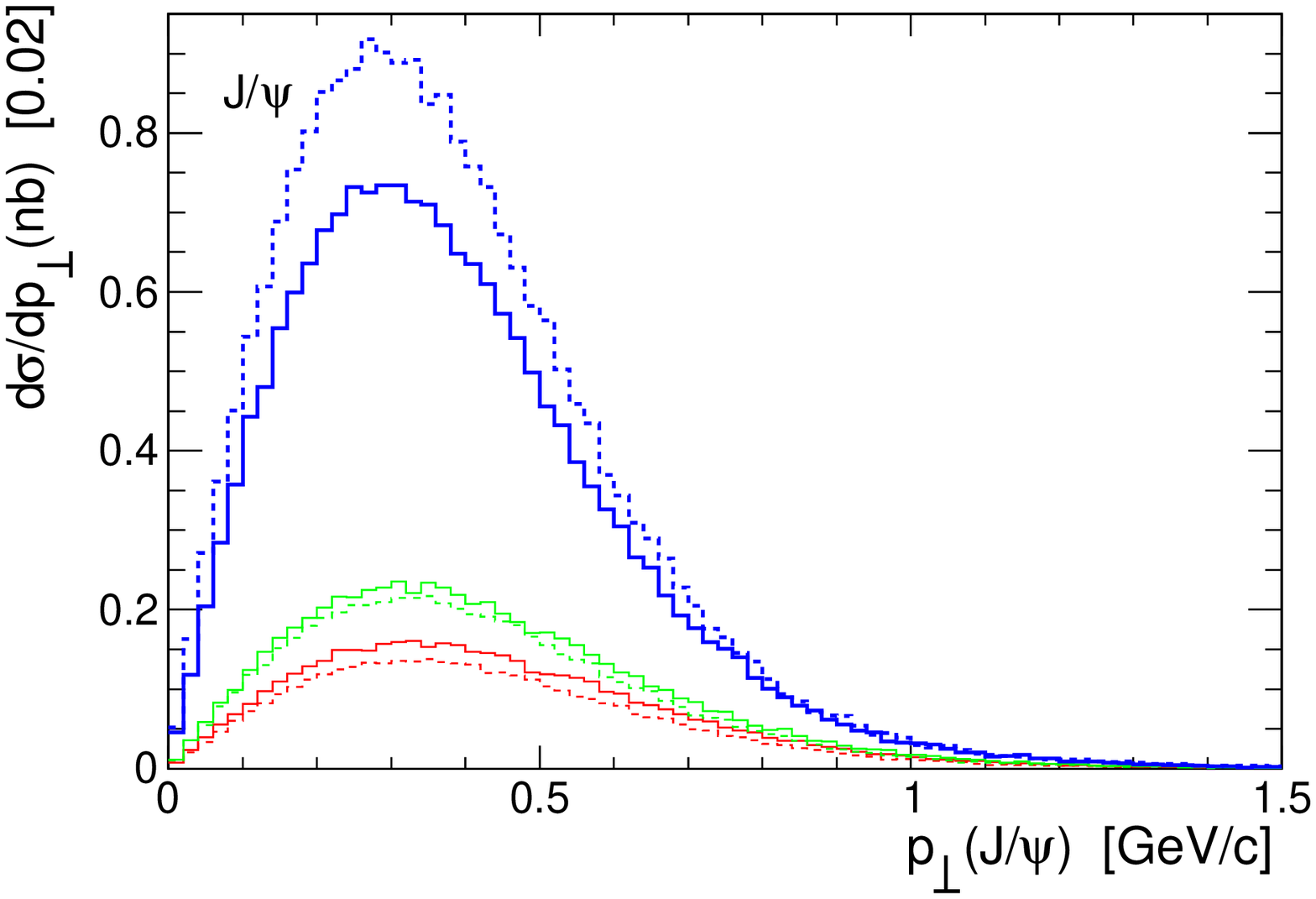}
\includegraphics[width=0.45\textwidth]{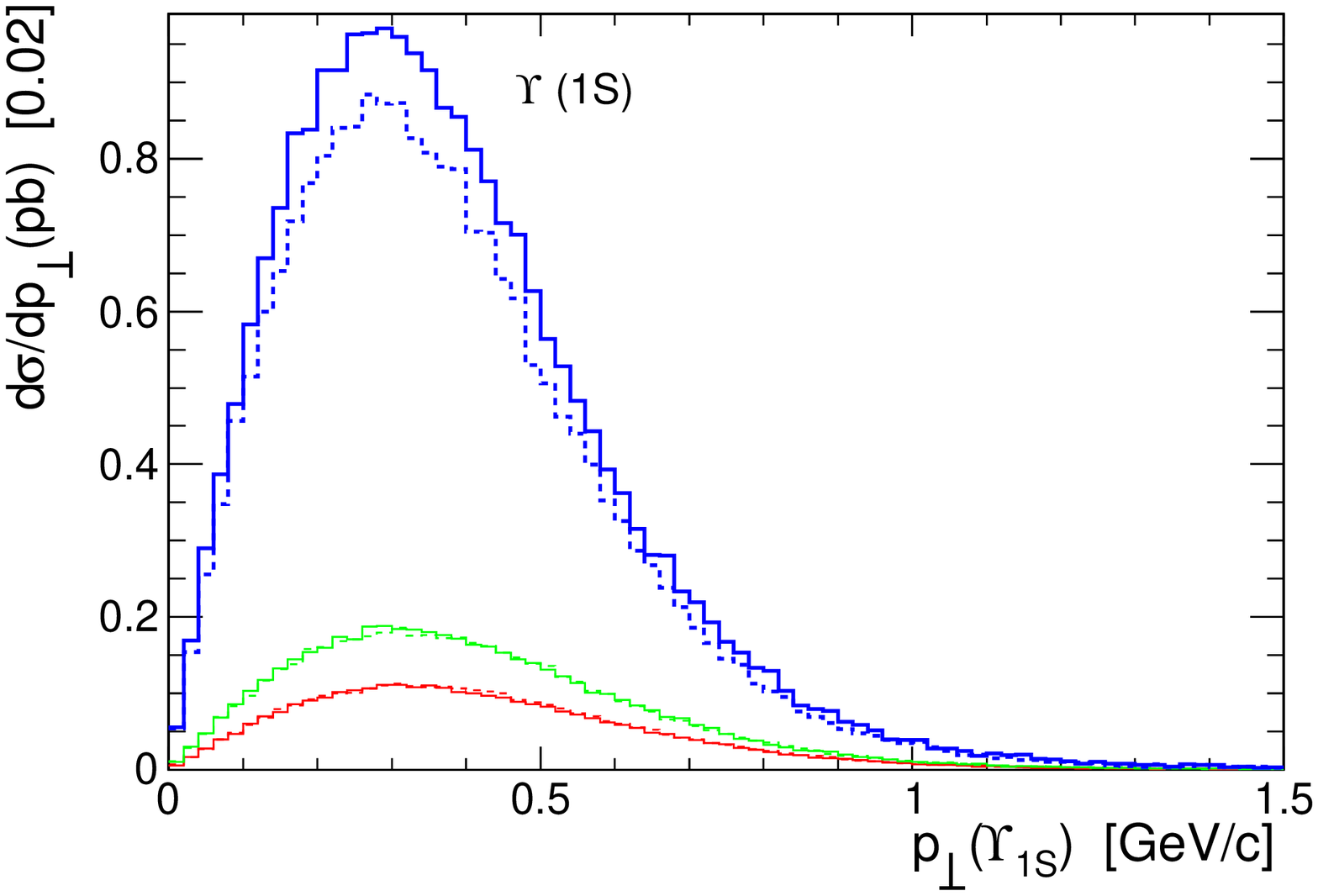}
\caption{\label{fig-yV-ptV}
Rapidity (upper panels) and transverse momentum (lower panels) distributions at LHC and FCC energy regimes  for the $J/\Psi$ (left) and $\Upsilon$ (right) photoproduction. }
\end{figure}
\end{center}

Lets now analyse the final-state distributions. In Fig.~\ref{fig-yV-ptV} we present our predictions for the rapidity (upper panels) and transverse momentum (lower panels) distributions for different values of the c.m. energy.  This is a useful result, since it can be directly obtained from the phenomenological models, as well as from the analysis of the experimental results, as seen in Refs.~\cite{lhcb2,lhcb3}.
Considering initially the rapidity distribution for the $J/\Psi$ production (upper left panel), one have that the default and bCGC predictions are very similar in the forward and backward regions at 7 and 13~TeV, but differ at mid-rapidities. In contrast, at FCC energies, they are similar at mid-rapidities and differ by $\approx$30\% for $|Y|$~=~5. In the case of the $\Upsilon$ production, we have that the difference between the predictions is smaller. 
In addition, we have analysed the transverse momentum distribution of the vector mesons, as shown in the bottom panel of Fig.~\ref{fig-yV-ptV}. The differences are more significant at 100~TeV, especially for the $J/\Psi$ meson. Another observation in these results is the slightly shift of the transverse momentum distributions towards smaller values of $p^{V}_{\perp}$ when the energy increases, which is expected from Eq.~(\ref{bv}).

Looking at the distributions of the muons coming from the decay of the vector mesons, applying specific kinematical cuts that may be useful in experimental analyses of exclusive events. We focus on muons in this work given that the LHC experiments are highly efficient in their detection in hadronic collisions \cite{cmstdr,atlastdr}. In this case, one has to apply proper kinematical cuts to reduce the contamination from the inclusive background -- i.e. Drell-Yan dimuon production -- in order to enhance the signal region. Thus, for the current data-taking at the LHC, we focus this investigation on $pp$ collisions at 13~TeV, selecting the events by applying a set of kinematic cuts on the single muons, such as $\eta(\mu^{\pm})<$~2.5 and $p_{\perp}(\mu^{\pm})>$~0.5~GeV for the $J/\Psi$, whereas for the $\Upsilon$(1S) we apply $\eta(\mu^{\pm})<$~2.5 and $p_{\perp}(\mu^{\pm})>$~4.0~GeV.
Considering that the continuum background from the exclusive two-photon $\gamma\gamma\to\mu^{+}\mu^{-}$ production plays an important role in this kinematic range, we produce an event sample for this background and include it along with the predictions for the vector meson photoproduction to simulate a more realistic scenario. Together with the previous kinematic cuts, it is possible to improve the event selection by applying additional cuts on the dimuons to suppress the contribution from the continuum (non-resonant) background, like $\Delta p_{\perp}(\mu^{+}\mu^{-})>$~0.05~GeV, $\Delta\phi(\mu^{+}\mu^{-})>$~0.01, and $p_{\perp}(\mu^{+}\mu^{-})>$~0.04~GeV for the $J/\Psi$, and $\Delta p_{\perp}(\mu^{+}\mu^{-})>$~0.05~GeV, $\Delta\phi(\mu^{+}\mu^{-})>$~0.01, and $p_{\perp}(\mu^{+}\mu^{-})>$~0.12~GeV for the $\Upsilon$(1S) photoproduction.

\begin{center}
\begin{figure}[t!]
\hspace*{-3.5em}\includegraphics[width=0.45\textwidth]{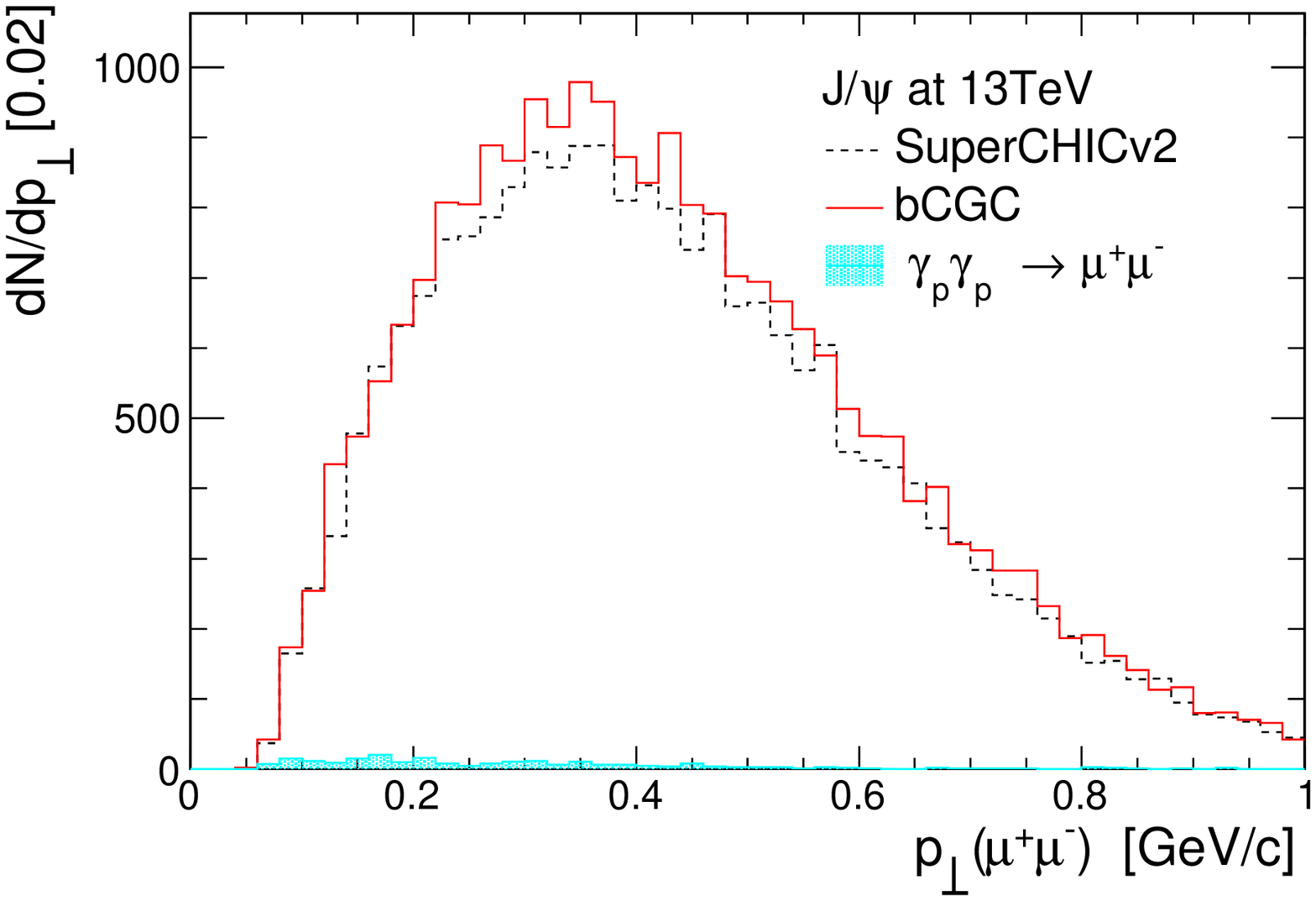}
\hspace*{-1.5em}\includegraphics[width=0.45\textwidth]{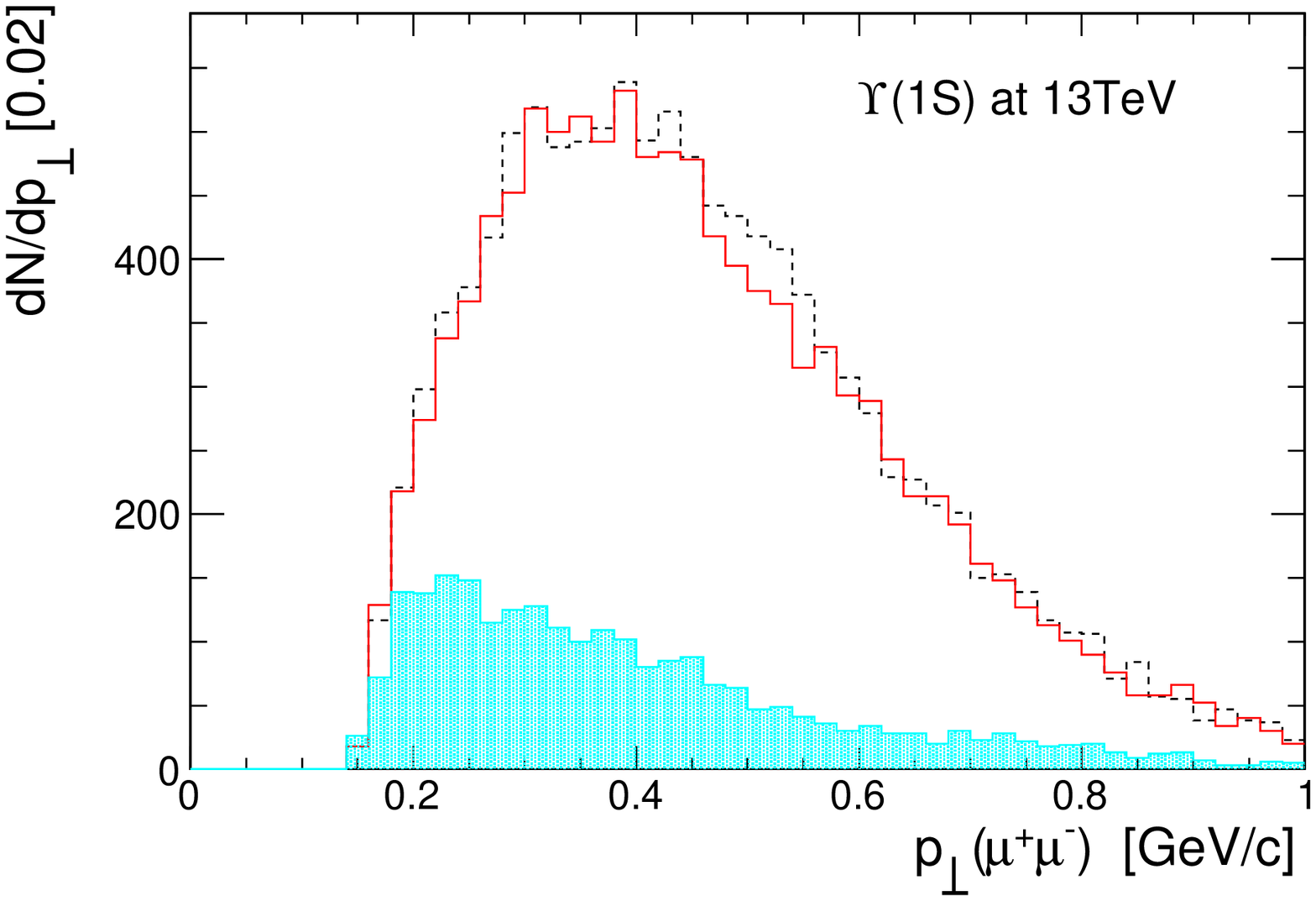}
\hspace*{-3.5em}\includegraphics[width=0.45\textwidth]{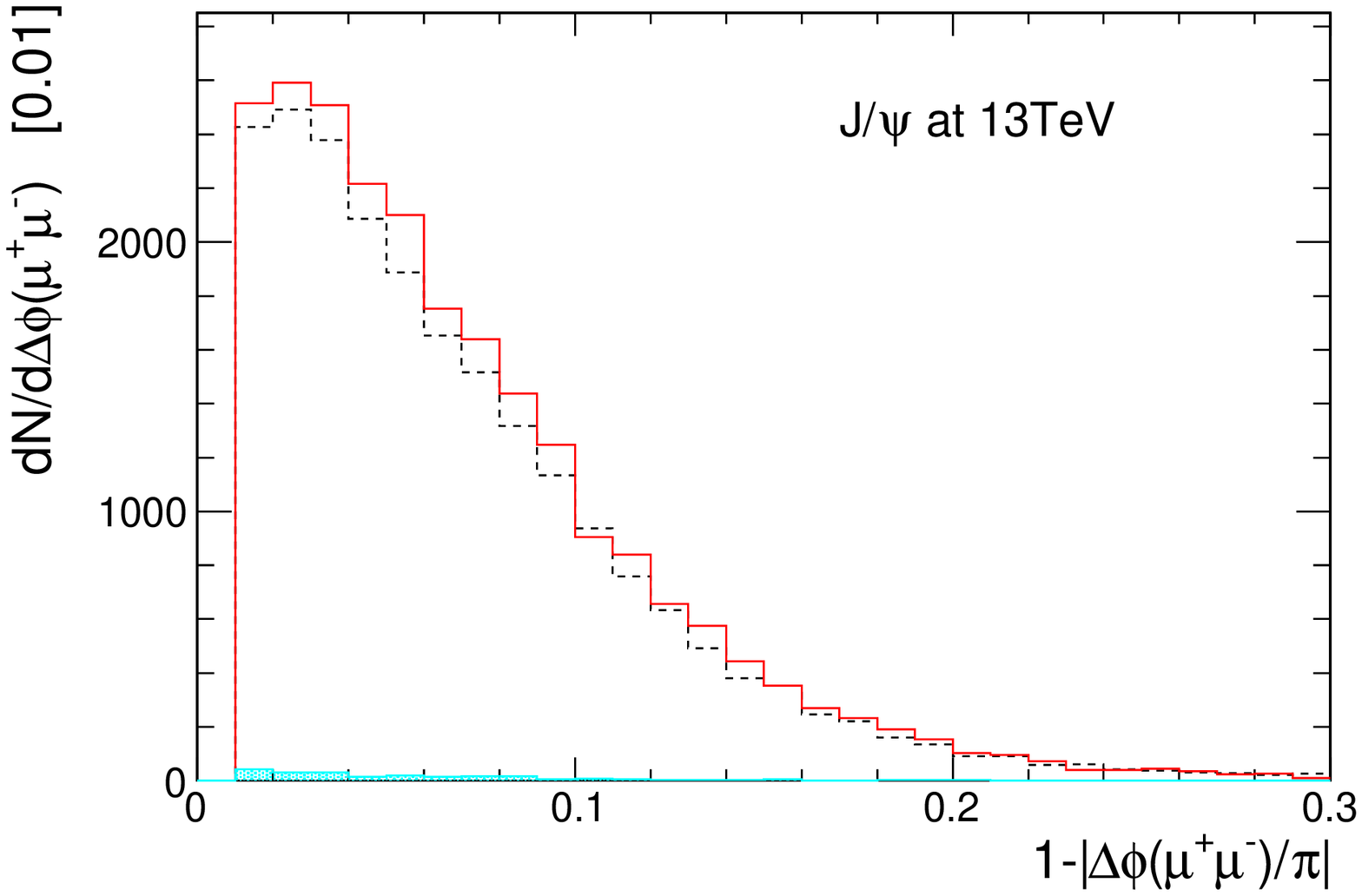}
\hspace*{-1.5em}\includegraphics[width=0.45\textwidth]{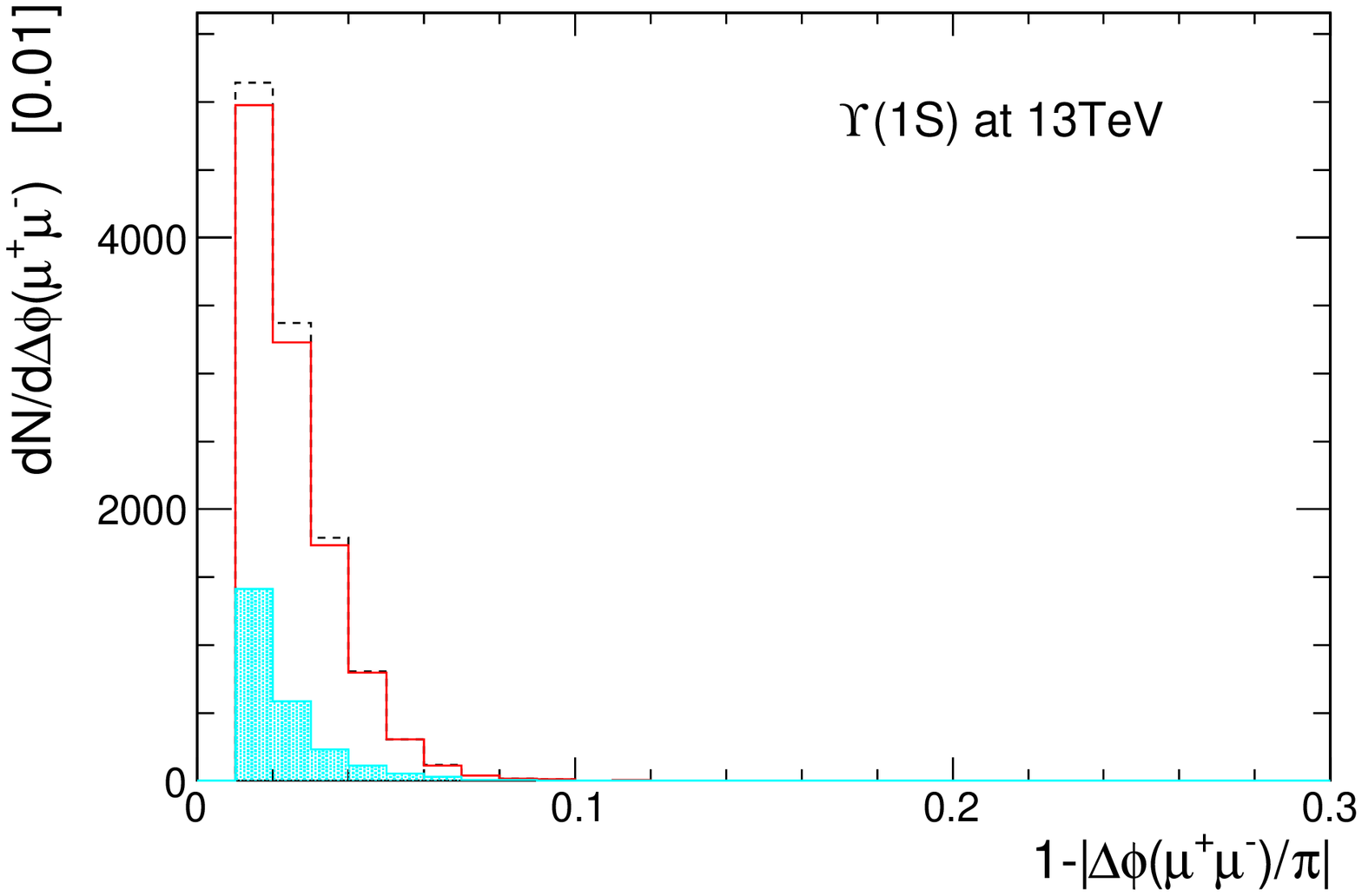}
\hspace*{-3.5em}\includegraphics[width=0.45\textwidth]{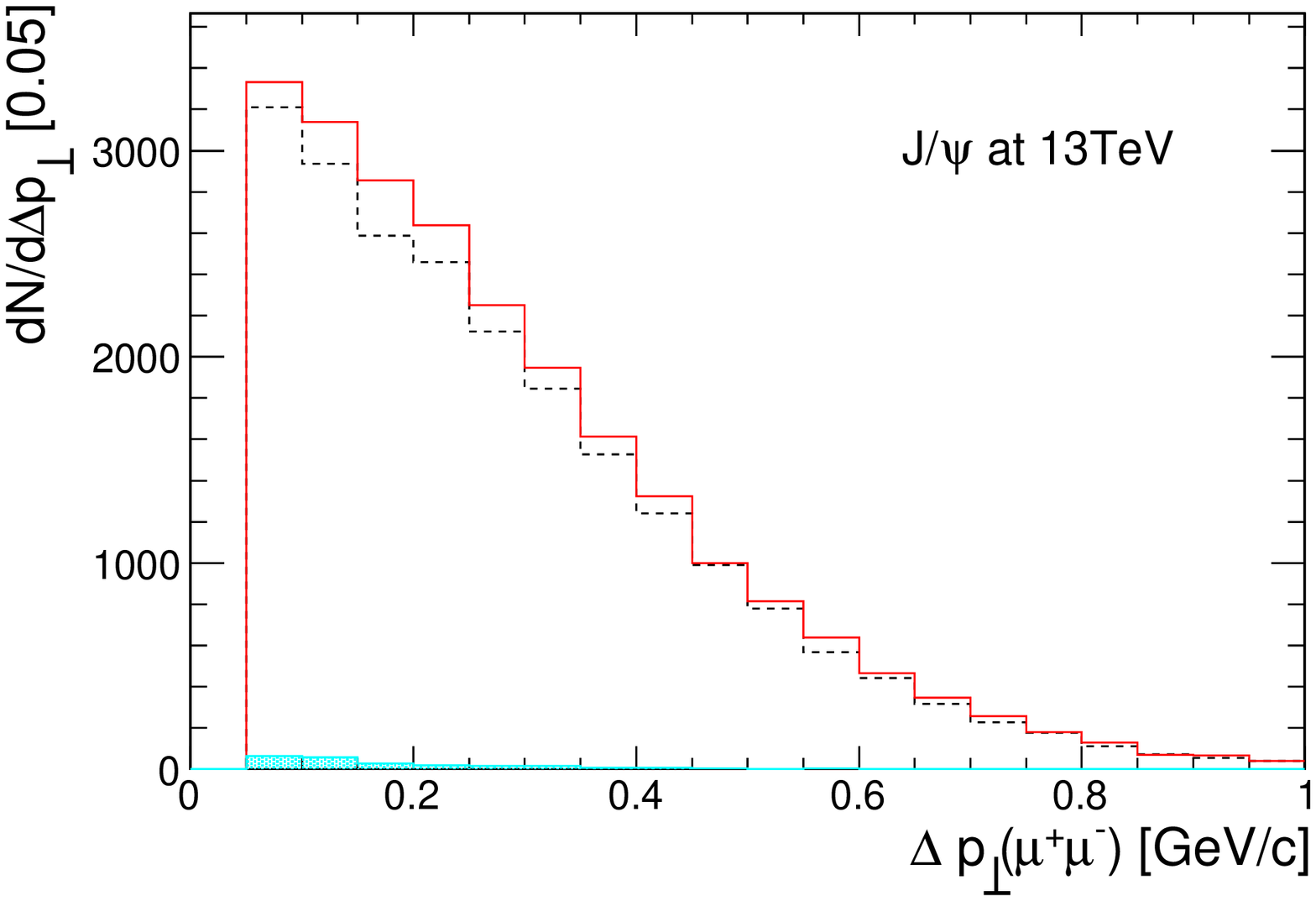}
\hspace*{-1.5em}\includegraphics[width=0.45\textwidth]{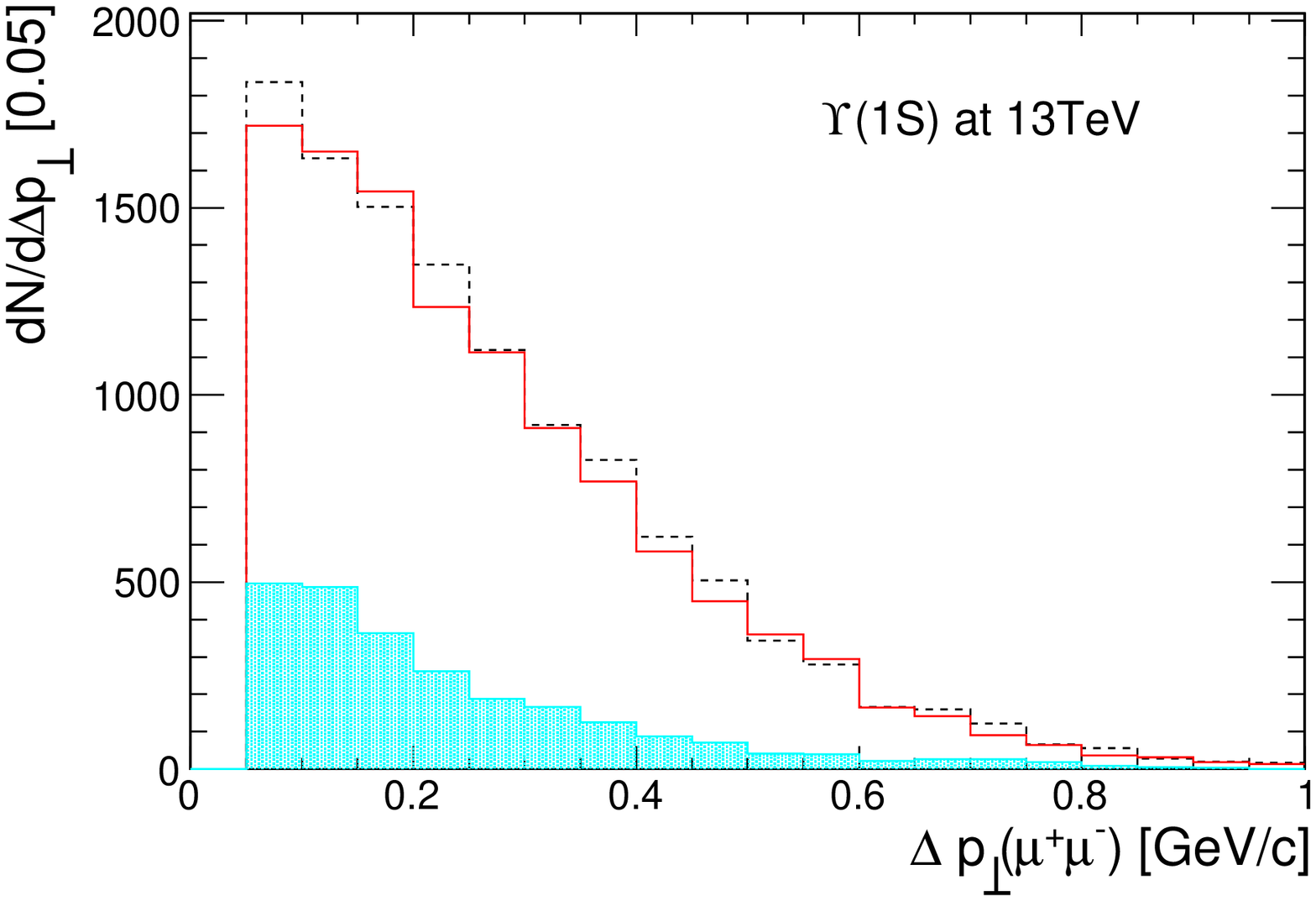}
\caption{\label{fig-acop-dpt}
Final-state distributions of the dimuons  from the decay of the $J/\Psi$ (left panels) and $\Upsilon$(1S) (right panels) in $pp$ collision at $\sqrt{s} = 13$ TeV.}
\end{figure}
\end{center}

\begin{center}
\begin{figure}[t!]
\hspace*{-3.5em}\includegraphics[width=0.45\textwidth]{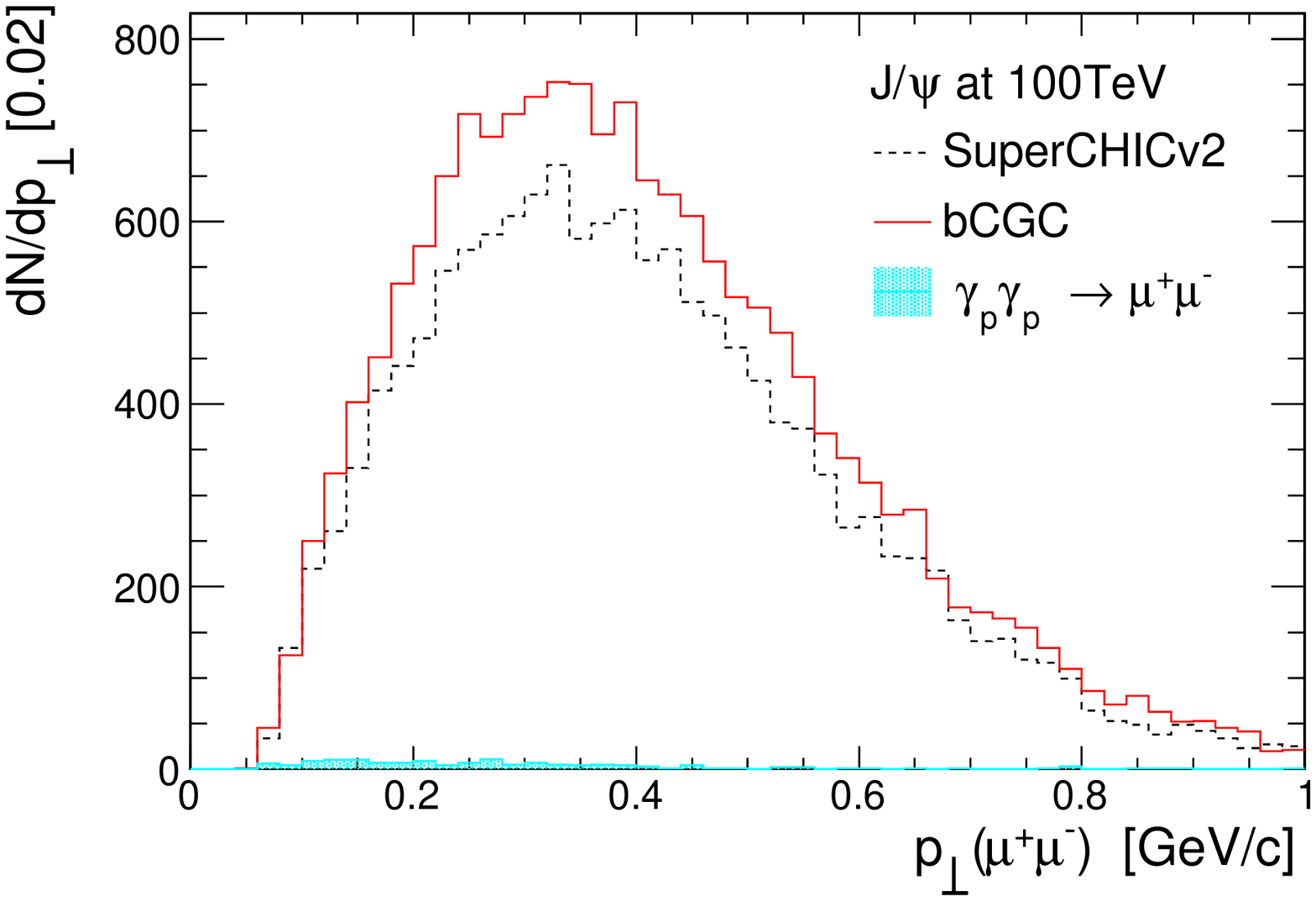}
\hspace*{-1.5em}\includegraphics[width=0.45\textwidth]{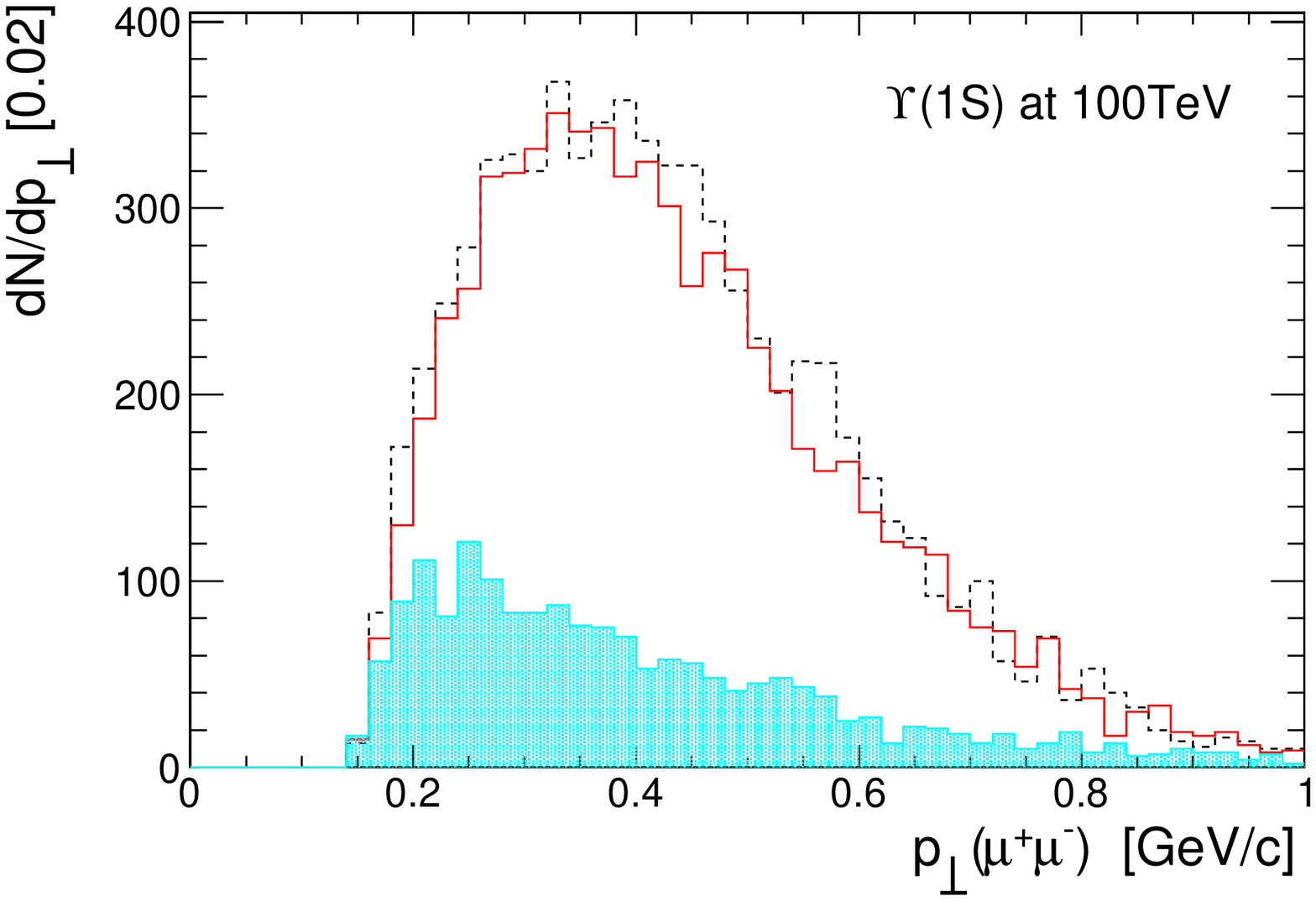}
\hspace*{-3.5em}\includegraphics[width=0.45\textwidth]{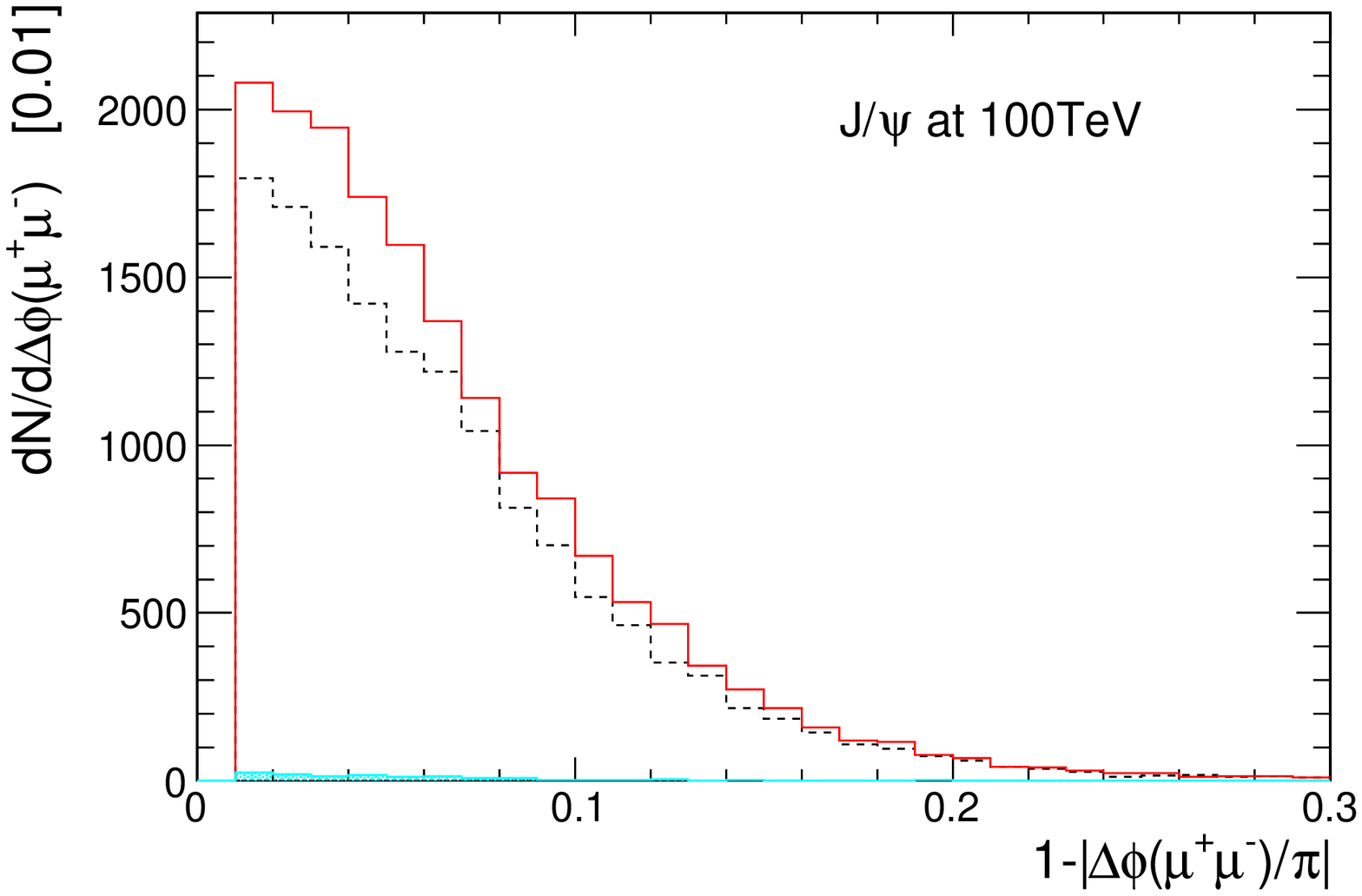}
\hspace*{-1.5em}\includegraphics[width=0.45\textwidth]{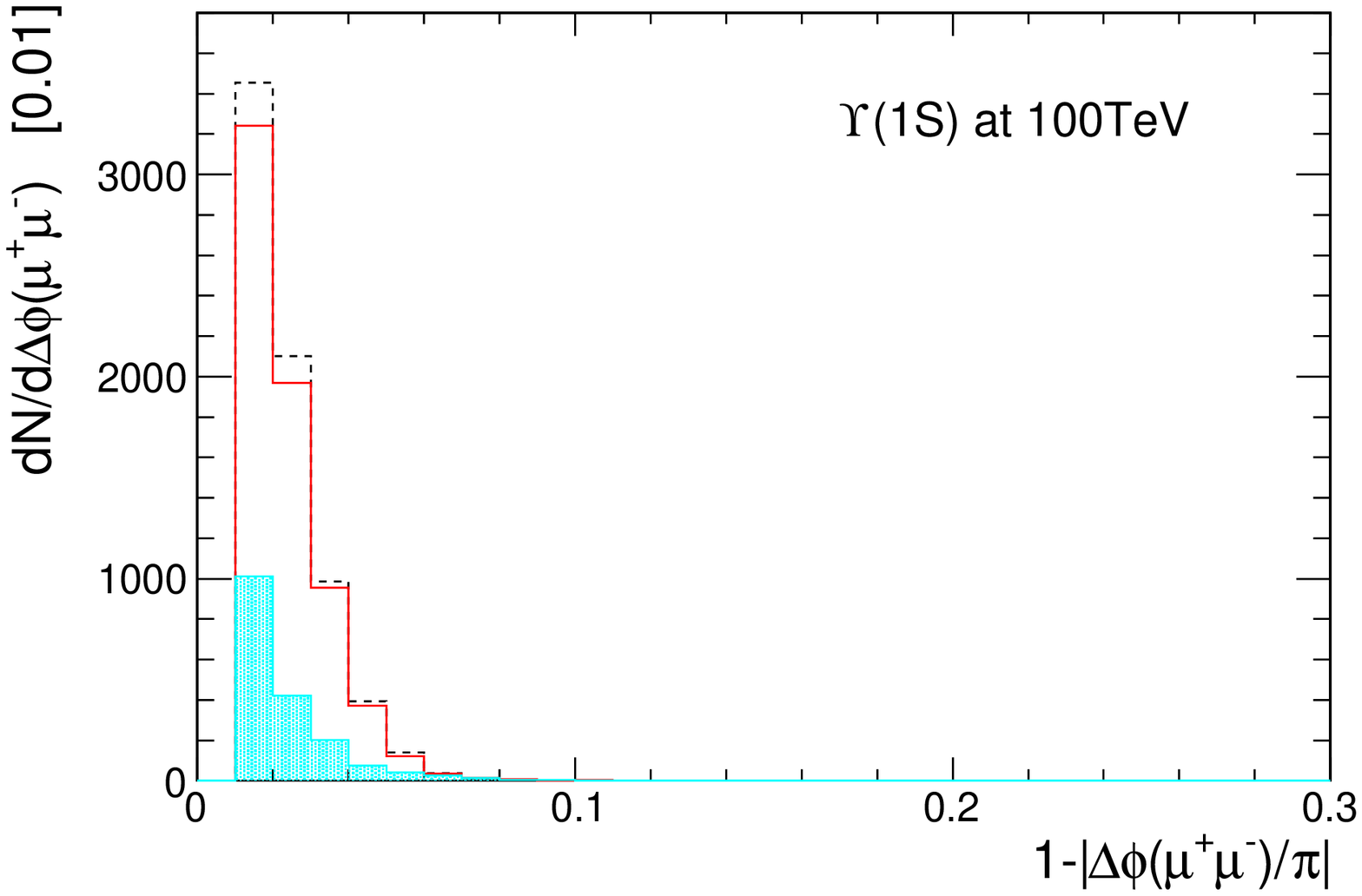}
\hspace*{-3.5em}\includegraphics[width=0.45\textwidth]{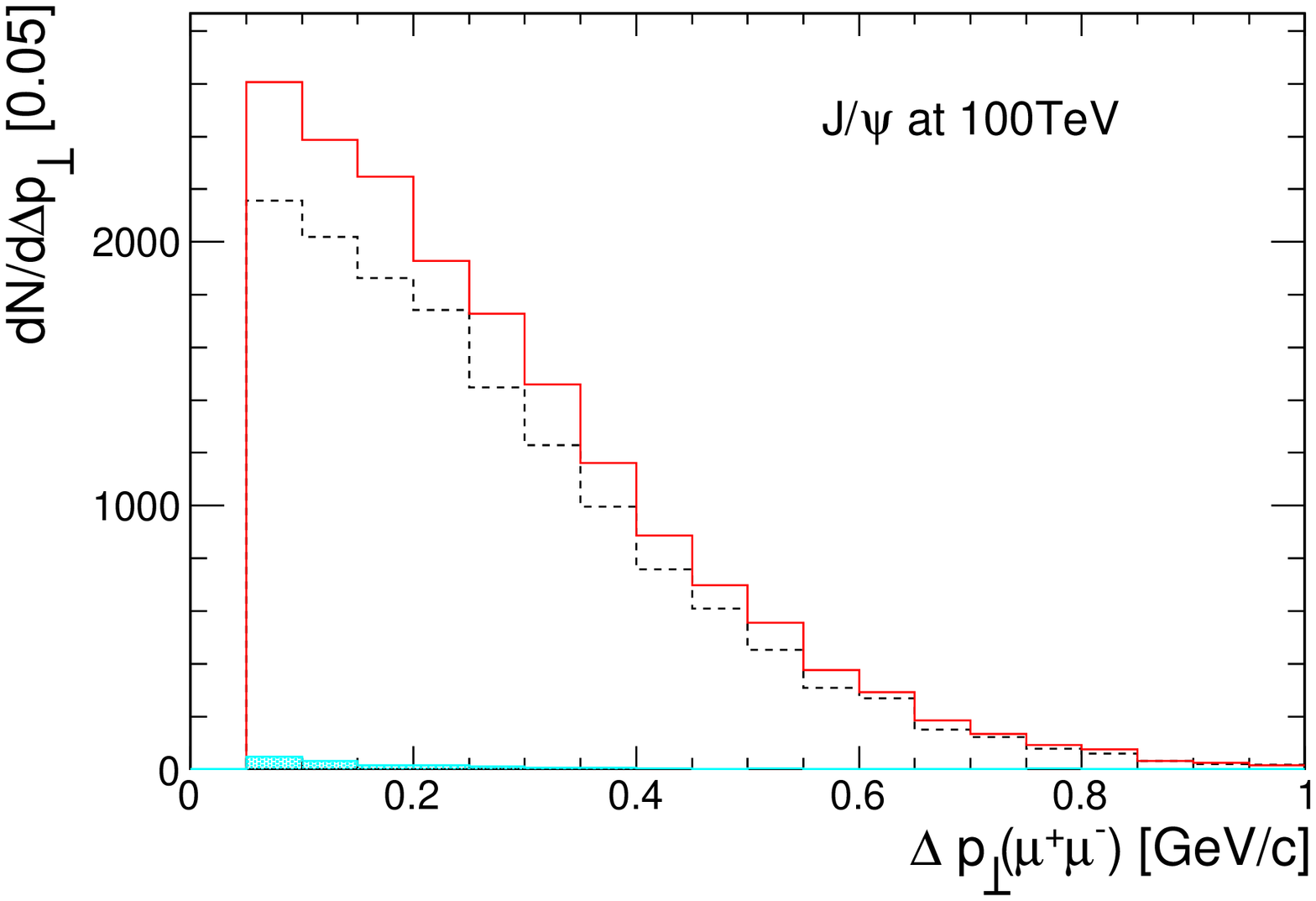}
\hspace*{-1.5em}\includegraphics[width=0.45\textwidth]{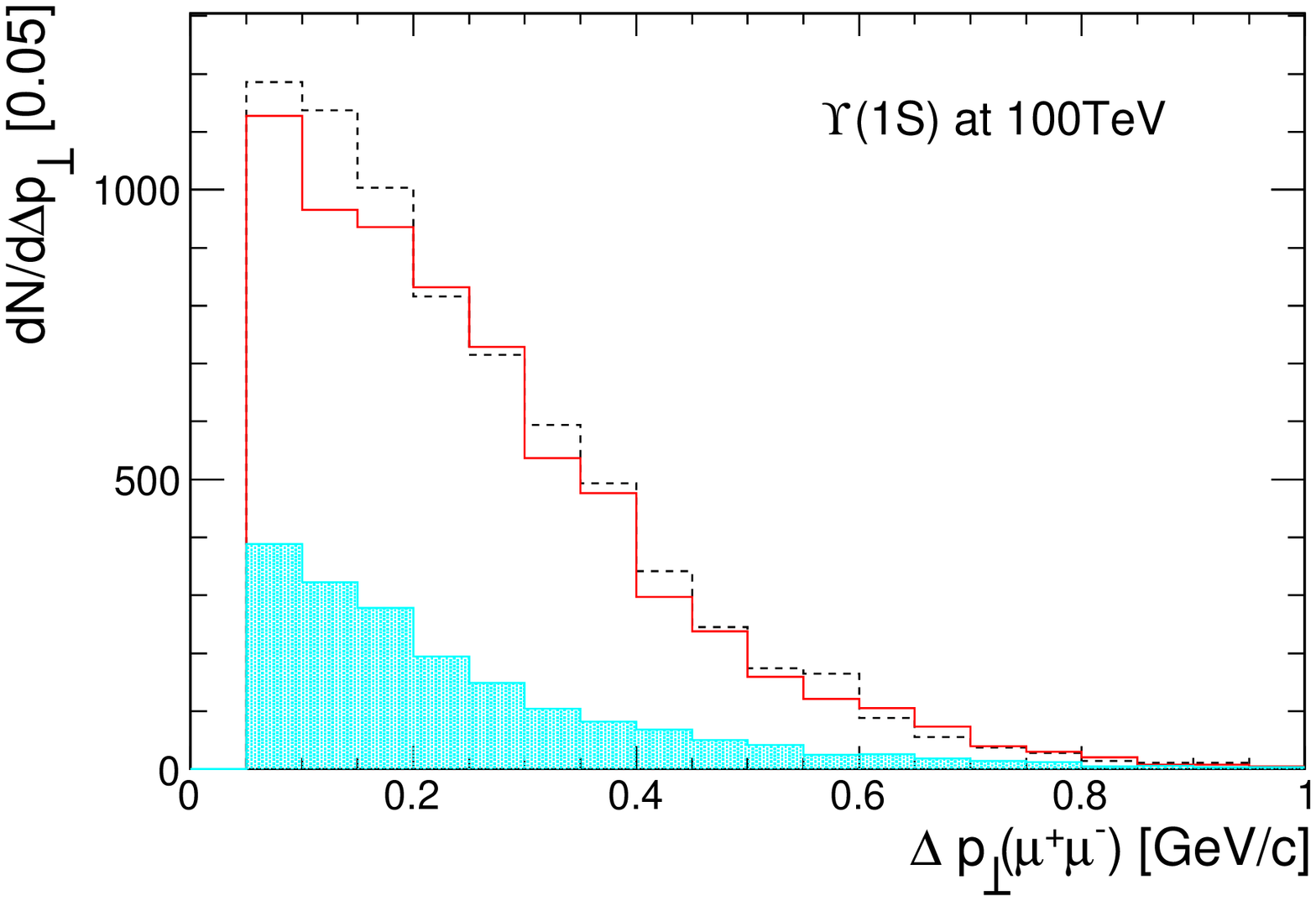}
\caption{\label{fig4}
Final state distributions of the dimuons  from the decay of the $J/\Psi$ (left panels) and $\Upsilon$(1S) (right panels) in $pp$ collision at $\sqrt{s} = 100$ TeV.}
\end{figure}
\end{center}

In Fig.~\ref{fig-acop-dpt} we present our predictions for  the transverse momentum distributions of the dimuons in Fig.~\ref{fig-acop-dpt}, with a distribution centred around $p_{\perp}(\mu^{+}\mu^{-})\simeq$~0.3~GeV.  Given that the two-photon production has its largest contribution from back-to-back muons, it has a steep fall from low values of dimuon transverse momentum \cite{fwd-10-005,fwd-11-004}. On the other hand, the vector meson photoproduction shows a distribution around larger values of $p_{\perp}(\mu^{+}\mu^{-})$ due to the  Pomeron exchange in the $t$-channel, and then the continuum background can be easily suppressed with kinematic cuts.
Figure~\ref{fig-acop-dpt} also shows the acoplanarity distributions of the dimuons at 13~TeV including the continuum background. It is clear that the difference between the default and bCGC predictions is more pronounced in the $J/\Psi$ case than in the $\Upsilon$(1S) one. Moreover, even the background from the two-photon production being more significant at low acoplanarity, the kinematical cuts applied can minimize its contribution in the $J/\Psi$ case, favouring a kinematic region for the comparison of the two models. Nevertheless, this region is totally contaminated in the $\Upsilon$(1S) case. Another kinematic variable of interest is $\Delta p_{\perp}(\mu^{+}\mu^{-})$, which measures the transverse momentum balance in the event, as shown in bottom panel of Fig.~\ref{fig-acop-dpt}. Once again, the $J/\Psi$ photoproduction may provide the best scenario for a comparison of the phenomenological models, while they are similar for the $\Upsilon$(1S) photoproduction. In both ranges, the contamination from inclusive background (such as Drell-Yan dimuon and resonant background) would be small or negligible, as shown in the reported analyses from the LHCb experiment \cite{lhcb3}. Then, the signal-to-background ($S/B$) ratio obtained within these set of kinematic cuts results in around 21 for the $J/\Psi$ photoproduction at the LHC at 13~TeV, showing the possibility of such study with the LHC data and comparison of the phenomenological models. 
Although the $\Upsilon$(1S) photoproduction can be also studied, the $S/B$ ratio goes down to 6.

The same study have been performed for the FCC energy regime, where the predictions show a more favourable scenario for a comparison of the phenomenological models. Figure~\ref{fig4} shows the kinematics distributions of the muon pairs in both $J/\Psi$ and $\Upsilon$(1S) photoproduction at 100~TeV. It is clear that the $J/\Psi$ is the best observable that allows a comparison among the theoretical predictions with the $\Upsilon$(1S) with nearly the same behaviour as in the LHC energy regime. Finally, the $S/B$ ratio is around twice larger than the results at 13~TeV, showing that the future data coming from FCC will allow a better tuning of the phenomenological models in comparison with the experimental data.


\section{Summary} 
\label{sec:sum}

The exclusive vector meson photoproduction in $pp$ collisions is an important probe of the QCD dynamics at high energies. Recent studies at Tevatron, RHIC and LHC have demonstrated that the experimental analysis of this process is feasible and that data can be used to constrain several aspects associated to its theoretical description. Such situation should be improved in a near future with the installation of forward detectors. In this paper we have analysed the exclusive $J/\Psi$ and $\Upsilon$ photoproduction in $pp$ collisions at the energies of the Run-II of LHC and presented predictions for the FCC for the first time. Our main emphasis was in the final state distributions that can be measured by the experimentalists. In order to estimated these distributions we have used the \texttt{SuperCHIC2} tuned by the LHCb data. Moreover, we have modified the energy dependence of the photon-hadron cross section in order to take into account the nonlinear effects in the QCD dynamics, as described by the bCGC model, and obtain more realistic predictions for higher energies that those probed in the Run-I. The impact of the modelling of the energy dependence of the $\gamma p \rightarrow V p$ cross section have been investigated. Finally, several final state distributions for the dimuons generated from the decay of the vector mesons were estimated considering realistic kinematic cuts and compared with  the background associated to the exclusive $\gamma \gamma \rightarrow \mu^+ \mu^-$ production. Our predictions can be directly compared with the Run-II data and shed light on the  exclusive vector meson photoproduction at the FCC.

\begin{acknowledgments}
 This research was supported by CNPq, CAPES and FAPERGS, Brazil. 
\end{acknowledgments}



\end{document}